\documentclass[aps,prb,reprint,superscriptaddress,nofootinbib,floatfix]{revtex4-2}

\usepackage{amsmath,amssymb,bm}
\usepackage{graphicx}
\usepackage{booktabs}
\usepackage{xcolor}
\usepackage{microtype}
\usepackage{setspace}
\usepackage{hyperref}
\usepackage{url}
\usepackage{placeins}
\hypersetup{colorlinks=true,citecolor=blue!55!black,linkcolor=blue!55!black,urlcolor=blue!55!black}
\graphicspath{{../results/}}

\newcommand{\D}{\mathcal D}
\newcommand{\Lio}{\mathcal L}
\newcommand{\dd}{\mathrm d}
\newcommand{\Tr}{\operatorname{Tr}}
\newcommand{\1}{\mathbb 1}
\newcommand{\ket}[1]{|#1\rangle}
\newcommand{\bra}[1]{\langle#1|}

\begin{document}

\title{Kardar--Parisi--Zhang density hydrodynamics in a coherent XXZ--ASEP Lindbladian}

\ifdefined\PRBpreview
\author{Yansheng Liu, Daoguang Li, Kaixiang Lu, Mingdi Xu, and
Lei Pan\thanks{Contact author: panlei@nankai.edu.cn}\\
\small School of Physics, Nankai University, Tianjin 300071, China}
\date{}
\else
\author{Yansheng Liu}
\affiliation{School of Physics, Nankai University, Tianjin 300071, China}

\author{Daoguang Li}
\affiliation{School of Physics, Nankai University, Tianjin 300071, China}

\author{Kaixiang Lu}
\affiliation{School of Physics, Nankai University, Tianjin 300071, China}

\author{Mingdi Xu}
\affiliation{School of Physics, Nankai University, Tianjin 300071, China}

\author{Lei Pan}
\email{panlei@nankai.edu.cn}
\affiliation{School of Physics, Nankai University, Tianjin 300071, China}
\fi

\newcommand{\manuscriptabstract}{%
Coherent interactions can generate quantum correlations without changing a
stationary density current.  Whether that current still determines
long-wavelength transport is a dynamical question.  We study an XXZ chain
with asymmetric Lindblad hopping, whose stationary current has the
asymmetric simple exclusion process (ASEP) form although its populations
do not evolve autonomously.  The current curvature and static
susceptibility fix the leading Kardar--Parisi--Zhang (KPZ) length and
amplitude without a fitted length or amplitude conversion factor.  The density-visible
Liouvillian spectrum, real-space spreading, and Fourier relaxation support
KPZ scaling at the finite coherent couplings studied.  Central and
intermediate stationary density correlations are consistent with the
Pr\"ahofer--Spohn profile at this microscopic scale, with resolved
finite-time corrections.  Reservoir-matched classical references quantify
boundary effects, while symmetric hopping recovers diffusion and reduced
asymmetry produces the expected crossover.  These results connect an exact
nonlinear current to the observed long-wavelength KPZ transport of the
physical density in an interacting open quantum chain away from the isotropic Heisenberg point.
}

\ifdefined\PRBpreview
\PRBpreviewfrontmatter{\manuscriptabstract}
\else
\begin{abstract}
\manuscriptabstract
\end{abstract}
\maketitle
\fi

\section{Introduction}

The Kardar--Parisi--Zhang (KPZ) universality class connects stochastic growth
to anomalous transport in one dimension
\cite{KPZ1986,HalpinHealy1995,Corwin2012,QuastelSpohn2015,Spohn1991,Spohn2014}.
For a stationary conserved density, its characteristic signatures are a
spreading length proportional to $t^{2/3}$ and a universal non-Gaussian
correlation profile \cite{PraehoferSpohn2004,SasamotoSpohn2010,FerrariSpohn2006}.
The asymmetric simple exclusion process (ASEP) provides a microscopic
realization: the curvature of its density-dependent current generates the
Burgers nonlinearity, and its slow relaxation rate scales as $N^{-3/2}$
\cite{GwaSpohn1992,Derrida1998,Mallick2015,ProlhacMallick2008}.
KPZ transport also occurs in quantum magnets, most prominently at the
isotropic Heisenberg point
\cite{Znidaric2011,Ljubotina2017,Ljubotina2019}.  There it is associated with
integrable quasiparticle dynamics and non-Abelian symmetry
\cite{Bulchandani2018,DeNardisGHD2019,Gopalakrishnan2019,DeNardis2019,
DeNardis2020,DupontMoore2020,Ilievski2021,DeNardis2021,Ye2022}, and has
motivated experiments with magnetic materials, cold atoms, and
superconducting qubits \cite{Scheie2021,Wei2022,Rosenberg2024}.
These mechanisms raise a different question for open quantum matter: can a
nonlinear stochastic current govern physical-density transport when
coherent interactions continuously generate quantum correlations?

Open quantum systems offer several settings in which to address this
question.  Quantum stochastic exclusion processes and extensions of
macroscopic fluctuation theory describe coherent fluctuations beyond
classical transport
\cite{BertiniMFT2015,Bernard2021,Eisler2011,Bauer2017,BernardJin2019,
BernardJin2021,BernardEssler2022,HruzaBernard2023,BarraquandBernard2025,
Bernard2026,Jin2020,RobertsonEssler2021}.  Exactly solvable Lindbladians
expose the structure of relaxation in particular operator sectors
\cite{Prosen2008,Medvedyeva2016,EsslerPiroli2020,deLeeuw2021,PencEssler2026},
while boundary-driven exclusion and XXZ chains connect coherent exchange
to particle transport \cite{Temme2012,Prosen2011}.
KPZ signatures have been found in a selected charge sector of the
integrable B3 model \cite{Wang2026}, in the spectral gap of a dissipative
Bose--Hubbard model \cite{Westhoff2026}, and in local Green functions of
noisy circuits whose charge transport remains diffusive \cite{McCulloch2026}.
These observables probe different relaxation sectors; the stationary
two-time correlation must be tested to establish KPZ hydrodynamics of the
physical density.

Here the stochastic and coherent processes move the same conserved charge.
The model combines XXZ exchange with local asymmetric hopping jumps and
belongs to the family analyzed by Ampelogiannis and Doyon
\cite{AmpelogiannisDoyon2026}.  Their diffusion bound and nonlinear-current
argument motivate a KPZ expectation, but do not establish its dynamical
exponent or scaling function.  The stationary current has exactly the
ASEP form at arbitrary coherent coupling, providing the Burgers
nonlinearity without an adjustable transport parameter.  Nevertheless,
generic diagonal states develop coherences, and their populations do not
evolve autonomously.  The dynamical issue is therefore whether these
coherences merely change short-scale relaxation or introduce an additional
slow field that invalidates the one-component description.  This requires both a spectral and a correlation test.  The slowest operator
mode need not carry density, and an apparent superdiffusive width may
reflect a crossover.  The normalized profile and its microscopic scale
provide further constraints that cannot be supplied by a width exponent
alone.

We address these questions through the Liouvillian spectrum and stationary
density dynamics.  At anisotropy $\Delta=0.7$, away from the isotropic
Heisenberg point, the density-visible slow branch, real-space broadening,
and Fourier relaxation consistently support KPZ transport.  The exact
current curvature and susceptibility also fix the leading length and
amplitude of the Pr\"ahofer--Spohn profile, allowing a comparison without
rescaling the data by a fitted length.  Reservoir-matched classical
references quantify finite-geometry effects; symmetric jumps recover
diffusion, and varying the asymmetry tests the predicted crossover.
The main long-time data use $J=0.2$, with a separate ensemble at $J=0.6$
testing the effect of stronger coherent motion.  These tests connect the exact stationary current to density transport
over the finite range of sizes, times, and couplings studied.

\section{Coherent XXZ--ASEP Lindbladian}
\label{sec:model}

ASEP transports particles by stochastic hops with a preferred direction;
XXZ exchange transports the same charge coherently.  Combining the two
processes allows the strength of coherent motion to be varied while the
stationary density distribution remains fixed.  We first formulate the
bulk model on a periodic chain of $N$ sites.  The reservoir-coupled geometry
used for real-time correlations is introduced in Sec.~\ref{sec:method}.
In the hard-core-boson representation, $b_j$ annihilates a particle and
$n_j=b_j^\dagger b_j$ is the local density, with occupations 0 and 1.  The Lindblad equation \cite{Gorini1976,Lindblad1976} is
\begin{align}
 \dot\rho={}&\Lio(\rho)=-i[H_{\rm XXZ},\rho]
 \nonumber\\[-2pt]
 &+\sum_j\D[L^R_j](\rho)+\sum_j\D[L^L_j](\rho),
 \label{eq:lindblad}\\
 \D[L](\rho)={}&L\rho L^\dagger-\tfrac12L^\dagger L\rho
 \nonumber\\[-2pt]
 &-\tfrac12\rho L^\dagger L,
 \label{eq:dissipator}
\end{align}
where
\begin{align}
H_{\rm XXZ}={}&J\sum_j\bigl[b_{j+1}^\dagger b_j+b_j^\dagger b_{j+1}
 \nonumber\\[-2pt]
&\hspace{2.5em}+2\Delta(n_j-\tfrac12)(n_{j+1}-\tfrac12)\bigr],
 \label{eq:xxz}\\
L^R_j={}&\sqrt p\,b_{j+1}^\dagger b_j,\qquad
L^L_j=\sqrt q\,b_j^\dagger b_{j+1}.
\label{eq:jumps}
\end{align}
We use $\hbar=1$ and unit lattice spacing, with $p,q\ge0$, $p+q>0$, and real $J,\Delta$.
Time is measured in units of the inverse mean hopping rate,
$(p+q)/2=1$.  Unless stated otherwise we take
$J=0.2$, $\Delta=0.7$, $p=2$, $q=0$, and half filling.

The total charge $Q=\sum_jn_j$ commutes separately with $H_{\rm XXZ}$ and
every jump operator.  Thus $U(1)$ is a \emph{strong} Lindbladian symmetry
\cite{BucaProsen2012,AlbertJiang2014}: fixed-$Q$ density operators form
invariant sectors.
At $J=0$ the diagonal operator subspace evolves as ASEP.  It ceases to be
invariant for $J\ne0$.  Indeed, for a diagonal state
$\rho_{\rm d}=\sum_sP_s\ket{s}\bra{s}$,
\begin{equation}
 \bra{s}[-iH_{\rm XXZ},\rho_{\rm d}]\ket{s'}
 =-i(H_{\rm XXZ})_{ss'}(P_{s'}-P_s),
 \label{eq:diagonal-leakage}
\end{equation}
which is nonzero whenever two configurations connected by the XX exchange have
different probabilities.  Thus a generic diagonal state does not obey a
closed classical rate equation once $J$ is finite.

\subsection{Stationary family and exact current}
\label{sec:current}

Let $\rho_M=\1_M/\dim\mathcal H_M$ be the normalized identity in the sector
with $M$ particles.  It is stationary for arbitrary $J$ and $\Delta$.
The Hamiltonian part annihilates $\1_M$, while
\[
 \sum_j\D[L^R_j](\1_M)=p\sum_j(n_{j+1}-n_j)\1_M=0,
\]
and similarly for the left jumps.  Equivalently, the grand-canonical product
family
\begin{equation}
 \rho(\bar n)=\bigotimes_j\left[(1-\bar n)\ket0\bra0
 +\bar n\ket1\bra1\right]
 \label{eq:product-steady}
\end{equation}
is stationary.

Taking the positive direction from $j$ to $j+1$, the continuity equation
$\Lio^\dagger(n_j)=\hat\jmath_{j-1}-\hat\jmath_j$ gives the microscopic current
\begin{align}
 \hat\jmath_j={}&-iJ(b_{j+1}^\dagger b_j-b_j^\dagger b_{j+1})
 \nonumber\\[-2pt]
 &+pn_j(1-n_{j+1})-q(1-n_j)n_{j+1}.
 \label{eq:current-op}
\end{align}
The coherent contribution is traceless in $\mathcal H_M$.  Both oriented
particle--hole configurations occur in $\rho_M$ with probability
$M(N-M)/[N(N-1)]$.  Hence
\begin{equation}
 j_N(M)=(p-q)\frac{M(N-M)}{N(N-1)}.
 \label{eq:finite-current}
\end{equation}
Taking $M/N\to\bar n$ gives
\begin{equation}
 j(\bar n)=(p-q)\bar n(1-\bar n),\qquad
 j''(1/2)=-2(p-q).
 \label{eq:thermo-current}
\end{equation}
The finite-size result and its thermodynamic limit are independent of $J$ and
$\Delta$.  The bulk relation is a specialization of the current obtained
for a broader family of charge-conserving Lindbladians
\cite{AmpelogiannisDoyon2026}.  ED at $N=6$ for all $M=1,\ldots,5$ and several values of $J$
reproduces Eq.~\eqref{eq:finite-current} to machine precision;
Eq.~\eqref{eq:thermo-current} then follows analytically as $N\to\infty$.
The finite-sector dependence is illustrated in Fig.~\ref{fig:current}; the independent small-system checks are summarized in Appendix~\ref{app:numerics}.

\begin{figure}[t]
 \centering
 \includegraphics[width=\columnwidth]{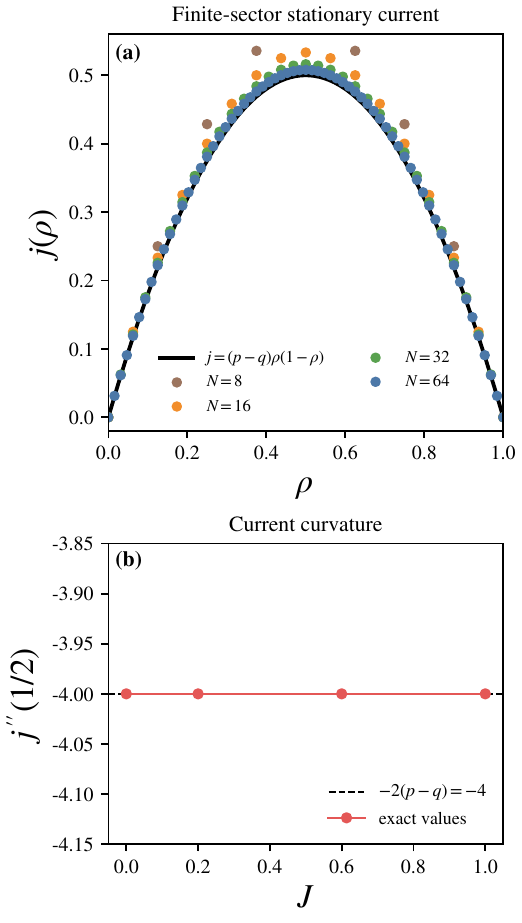}
 \caption{\label{fig:current}Stationary current and its curvature.
 (a) Finite-size sector averages, Eq.~\eqref{eq:finite-current}, approach
 $j(\bar n)=(p-q)\bar n(1-\bar n)$.  Symbols evaluate the exact
 finite-sector expression.
 Panel (b) shows that, for $p=2$ and $q=0$, the curvature
 $j''(1/2)=-4$ is independent of the coherent coupling; the horizontal line
 is the exact thermodynamic result.}
\end{figure}

\subsection{Nonlinear fluctuating hydrodynamics}

Suppose that local equilibration holds and that charge is the only conserved
field with a diverging relaxation time
\cite{Spohn2014,vanBeijeren2012}.  At half filling the mean drift
vanishes, $j'(1/2)=0$.  Expanding the fluctuating continuity equation for
$u(x,t)=n(x,t)-1/2$ gives
\begin{equation}
 \partial_tu+\frac{j''(1/2)}2\partial_xu^2
 =D\partial_x^2u+\partial_x(\sqrt\sigma\,\eta)+\cdots,
 \label{eq:burgers}
\end{equation}
Here $\eta$ has zero mean and covariance
$\langle\eta(x,t)\eta(x',t')\rangle=\delta(x-x')\delta(t-t')$.
Stationarity with susceptibility $\chi$ requires $\sigma=2D\chi$ in this
scalar convention; $D>0$ is nonuniversal.  For $p\ne q$, Eq.~\eqref{eq:burgers} is the noisy Burgers equation.
We define the stationary correlator in the product family by
$C(r,t)=\Tr[\delta n_r e^{t\Lio}(\delta n_0\rho(\bar n))]$,
with $\delta n_j=n_j-\bar n$.  In the periodic system,
$\sum_r C(r,t)=\chi$.  This grand-canonical normalization differs from the
zero sum enforced at fixed total charge; the fixed-charge ED below probes
nonzero-momentum decay rates.  At half filling, the predicted bulk scaling form is
\begin{equation}
 C(r,t)=\langle \delta n_r(t)\delta n_0(0)\rangle
 \simeq \chi(\Gamma t)^{-2/3}f_{\rm KPZ}
 \!\left[r/(\Gamma t)^{2/3}\right],
 \label{eq:kpz-profile}
\end{equation}
where the static susceptibility is $\chi=\bar n(1-\bar n)=1/4$ at half
filling.  We use the original unit-area Pr\"ahofer--Spohn function
\cite{PraehoferSpohn2004}, with $\int f_{\rm KPZ}(s)\,\dd s=1$ and
$\mu_2=\int s^2 f_{\rm KPZ}(s)\,\dd s\simeq0.510523$.
The second moment fixes the horizontal convention as well as the unit area.
Writing $u=\sqrt\chi\,\phi$ gives the nonlinear coupling
$G=j''(\bar n)\sqrt\chi/2$ in the unit-susceptibility field.  In this convention
the one-component prediction is \cite{Spohn2014}
\begin{equation}
 \Gamma=2\sqrt2\,|G|=\sqrt{2\chi}\,|j''(\bar n)|,
 \qquad \ell(t)=(\Gamma t)^{2/3}.
 \label{eq:kpz-amplitude}
\end{equation}
Thus $p=2$, $q=0$, and half filling give $\Gamma=2\sqrt2$ and
$\ell(t)=2t^{2/3}$.  The quantum model and TASEP therefore
share the leading scale at matched $p$, $q$, and $\bar n$, although their
finite-time corrections can differ.  Here $\Gamma$ denotes the density
scaling rate in the stated convention.  When $p=q$, the quadratic term
vanishes and this description predicts diffusion.

The approach to KPZ can be tested independently by reducing the asymmetry.
Writing $p=1+\epsilon$ and $q=1-\epsilon$, the Burgers nonlinearity is relevant
at the Edwards--Wilkinson fixed point and produces the crossover scales
\begin{equation}
 \ell_\times\sim\epsilon^{-2},\qquad
 t_\times\sim\epsilon^{-4},
 \label{eq:crossover-scales}
\end{equation}
consistent with the weakly asymmetric exclusion-process limit
\cite{BertiniGiacomin1997,GoncalvesJara2014},
and the weak-asymmetry scaling hypothesis gives
\begin{equation}
 4C(r,t;\epsilon)\simeq\epsilon^2\mathcal F
 (\epsilon^2r,\epsilon^4t).
 \label{eq:crossover-form}
\end{equation}
This is a weak-asymmetry scaling form at fixed $J$ and $\Delta$;
$\mathcal F$ retains their nonuniversal metric factors.  It presumes finite
short-scale transport coefficients as $\epsilon\to0$.

\section{Slow modes and density visibility}
\label{sec:spectrum}

The nonlinear current suggests a one-component hydrodynamic theory, but
this interpretation also requires charge to be the only slow conserved
field.  The jump algebra provides a first constraint: within a fixed-charge
sector, any operator commuting with every jump and its adjoint is
proportional to the identity.  Thus it supplies no additional exact local
conservation law beyond charge.  The argument and its restriction to the
charge-diagonal algebra are given in Appendix~\ref{app:commutant}.
An emergent slow coherence mode is still possible and must be tested in
the spectrum.

\begin{figure}[t]
 \centering
 \includegraphics[width=\columnwidth]{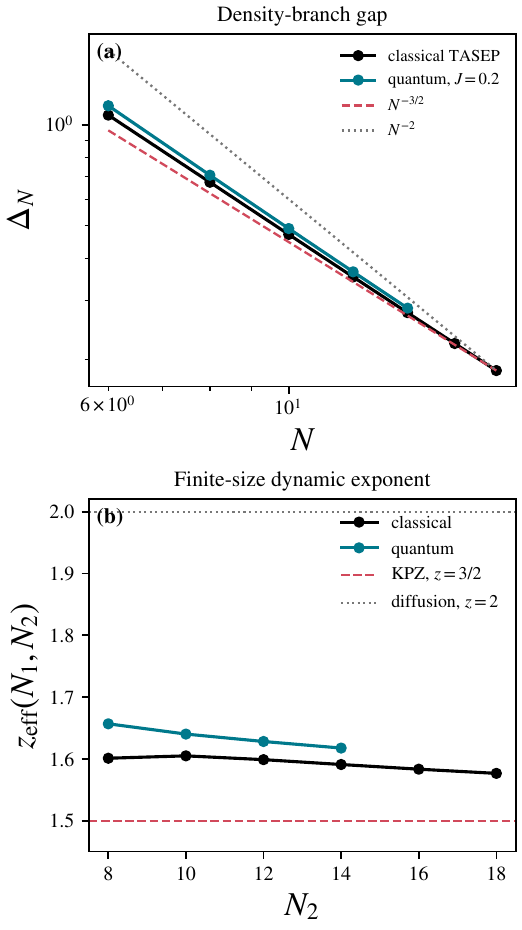}
 \caption{\label{fig:spectrum}Low-lying density modes at $J=0.2$.
 (a) Decay rates at $k=2\pi/N$, compared with TASEP and $N^{-3/2}$ and
 $N^{-2}$ guides anchored to the same largest-$N$ classical value.
 (b) Pairwise effective exponents, where $N_2$ is the larger member of each
 consecutive pair $(N_2-2,N_2)$.  Quantum data terminate at $N=14$, where the
 translation-resolved Liouville block already has dimension $841330$; the
 classical TASEP calculation extends to $N=18$.  Dashed and dotted horizontal
 lines mark $z=3/2$ and $z=2$, respectively.  Lines connecting data are guides
 to the eye; complementary finite-time scaling tests are given below.}
\end{figure}

Translation symmetry
allows the smallest nonzero momentum sector, $k=2\pi/N$, to be treated efficiently.  Figure~\ref{fig:spectrum}(a) shows the
density-branch decay rate
$\Delta_N=-\operatorname{Re}\lambda_{\rm dens}(2\pi/N)>0$
for $N=6$--$14$ at $J=0.2$.  Here $\lambda_{\rm dens}$ is the eigenvalue
of the density-visible branch; $\Delta_N$ differs from the anisotropy $\Delta$.
For $N_1<N_2$, we define
$z_{\rm eff}=\log(\Delta_{N_1}/\Delta_{N_2})/\log(N_2/N_1)$.
The pairwise effective
exponent decreases from $1.66$ to $1.62$ over these sizes, approaching the
TASEP trend and the KPZ value $3/2$.  The sizes are too small for an
independent asymptotic exponent, but they identify the branch that should
control the density response.

For $N\le12$, evolving a density modulation at $k=2\pi/N$ gives the same
decay rate as the slowest nonstationary mode in the fixed-$Q$ Liouvillian.
Thus the branch
in Fig.~\ref{fig:spectrum} is directly visible to the physical density and
does not require a fine-tuned initial-state projection.  The comparison with the full fixed-charge spectrum is available through
$N=12$; at $N=14$ the calculation is restricted to the density momentum block
(Appendix~\ref{app:numerics}).

Populations and coherences can be separated through an exact Schur complement.
On $\mathcal B(\mathcal H_M)$, let $P$ project onto occupation-diagonal
operators and $Q_o=\mathcal I-P$, where $\mathcal I$ is the identity
superoperator.  Write $\Lio=\Lio_0+J\mathcal V$, with
$\Lio_0=\Lio|_{J=0}$, $H_{\rm XXZ}=Jh_{\rm XXZ}$, and
$\mathcal V=-i[h_{\rm XXZ},\,\cdot\,]$; $h_{\rm XXZ}$ is independent of $J$.
Since
$P\mathcal VP=0$ and $\Lio_0$ preserves both subspaces, the effective
generator, wherever the coherence resolvent exists, is
\begin{equation}
 \Lio_{\rm eff}(\zeta)=P\Lio_0P
 +J^2P\mathcal VQ_o(\zeta-Q_o\Lio Q_o)^{-1}Q_o\mathcal VP.
 \label{eq:feshbach}
\end{equation}
Here $\zeta$ is a complex spectral parameter, distinct from the dynamical
exponent $z$; the inverse acts within the $Q_o$ subspace.
The frequency dependence in Eq.~\eqref{eq:feshbach} describes the memory
left by eliminating coherences.  If the coherence resolvent is regular,
the leading weak-coupling correction to population dynamics is of order
$J^2$; the resolvent itself retains its dependence on $J$.
At $J=0$, the tested off-diagonal decay rates within the half-filled,
charge-neutral sector remain separated from zero
while the density gap decreases with size.  The $J=0.2$ spectra are
consistent with the same separation over the accessible sizes.  This
supports the one-component description in the studied regime, without
establishing a uniform thermodynamic bound or the large-$J$ crossover time.

\section{Stationary density correlations}
\label{sec:method}

The real-time test concerns the spreading of a stationary density
fluctuation.  Long chains are treated with quantum-trajectory
matrix-product states \cite{Vidal2004,Dalibard1992,Daley2014,Weimer2021}.
Half-filled reservoirs maintain the product state $\rho_*=\1/2^N$:
injection and extraction rates are $p/2,q/2$ at the left end and
$q/2,p/2$ at the right end.  The connected correlation is
\begin{equation}
 C(r,t)=\overline{\langle\delta n_{y+r}(t)\rangle_{\mathcal T}
 \delta n_y(0)},
 \label{eq:estimator}
\end{equation}
where the average includes stationary initial configurations, trajectories,
and source positions in the central half of the chain.  The principal
ensemble has 512 trajectories at $N=512$ through $t=50$.
Direct evolution of the linear-response operator
$e^{t\Lio}(\delta n_y\rho_*)$ gives a complementary fixed-source
correlation without sampling noise.  Its size and source-position
dependence distinguish bulk spreading from the influence of the reservoirs.

A second observable measures relaxation across the full chain:
\begin{equation}
 S_N(k,t)=\frac1N\sum_{x,y=1}^N e^{-ik(x-y)}
 \Tr[\delta n_x e^{t\Lio}(\delta n_y\rho_*)],
 \label{eq:full-chain-structure}
\end{equation}
with $k=2\pi m/N$ and $S_N(k,0)=1/4$.  Figures use the shorthand $S=S_N$.
This definition includes boundary
sites as both sources and observation points.  It consequently has a
different boundary sensitivity from the central-source profile.
Although the product state is stationary, the reservoirs exchange charge
with the chain: the global conservation law and bulk correlation sum rule
are not exact in this geometry.  Classical comparisons therefore use both
periodic and reservoir-matched TASEP references.

The normalized structure factor is analyzed using the lattice wave number
$\widehat k=2\sin(|k|/2)$ and the variable $t\widehat k^z$.
For the collapse loss $\mathcal L(z)$, we minimize the mean squared
residual of $\operatorname{Re}S_N(k,t)/S_N(k,0)$ from a cubic polynomial in
$t\widehat k^z$, over the stated fit window.  Its four coefficients are
refitted at each $z$.  Complete trajectories are resampled to obtain statistical
intervals, repeating the normalization after ensemble averaging in each
resample.  The real-space and Fourier estimates share trajectories and
are complementary rather than statistically independent measurements.
A separate test compares the unadjusted profile with the absolute
prediction in Eq.~\eqref{eq:kpz-amplitude}; it fits neither a horizontal
scale nor an amplitude or background.  The central-window length used
below has an explicit mass constraint and serves only as a spreading
diagnostic.  Convergence and boundary controls are summarized in
Appendix~\ref{app:deterministic}.

\section{KPZ density transport}
\label{sec:longtime}

\begin{figure*}[t]
 \includegraphics[width=0.96\textwidth]{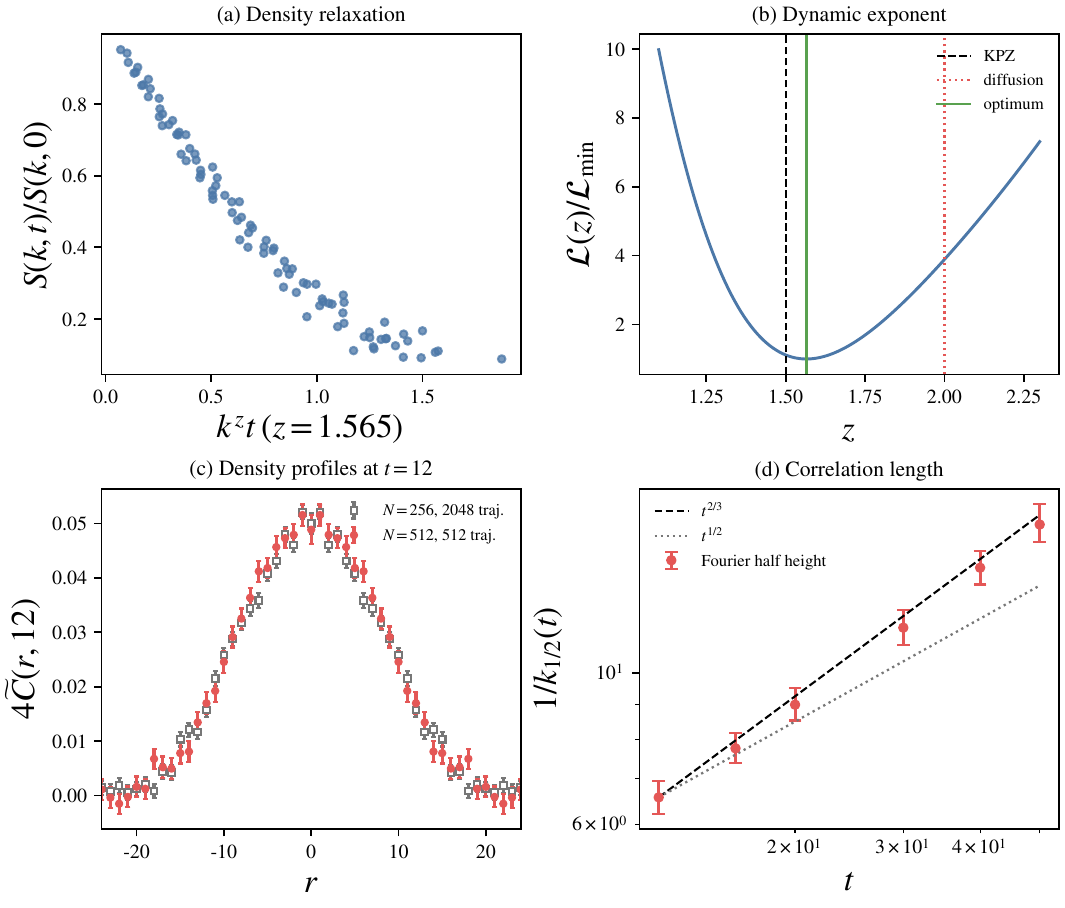}
 \caption{\label{fig:longtime}Density correlations for
 $J=0.2$, $\Delta=0.7$, $p=2$, and $q=0$.
 (a) Structure-factor collapse and (b) its loss, using the lattice wave number
 $\widehat k$ (denoted $k$ in the plot), on the broader diagnostic
 grid $8\le t\le50$, $m=4,\ldots,32$ at $N=512$.  The later-time,
 lower-wave-number calibration is shown separately in Fig.~\ref{fig:fourier}.  (c) Comparison of $N=256$ and
 $N=512$ profiles at $t=12$, with the window-mass convention defined in the text;
 the reduced $\chi^2$ is $0.88$.
 (d) Inverse Fourier half-height wave number of the mass-constrained
 central profile.  The $t^{2/3}$ and $t^{1/2}$ comparison guides are
 anchored at the same earliest-time value; neither is an additional fit.}
\end{figure*}

\subsection{Spreading of a density fluctuation}

A KPZ density mode spreads faster than a diffusive one.  The central-source
profiles in Fig.~\ref{fig:longtime} exhibit this broadening together with
an approximate collapse of the Fourier relaxation.  To quantify the
spreading without relying on a single spatial bin, we define a length
from the central profile at $N=512$.  On the window $W=\{-64,\ldots,64\}$, we symmetrize
$\bar C(r,t)=[C(r,t)+C(-r,t)]/2$ and remove a uniform offset:
$\widetilde C(r,t)=\bar C(r,t)-[\sum_{r'\in W}\bar C(r',t)-\chi]/129$.
Thus $\sum_{r\in W}\widetilde C(r,t)=\chi=1/4$.
For the continuous probe wave number $k\ge0$, define
$F_C(k,t)=\chi^{-1}\sum_{r\in W}\widetilde C(r,t)\cos(kr)$.
The first half-height crossing of its running-minimum envelope defines
$\ell_C(t)=1/k_{1/2}(t)$.
Over $12\le t\le50$,
\begin{equation}
 \ell_C(t)\sim t^{\beta_C},
 \label{eq:width-exp}
\end{equation}
with $\beta_C=0.649$ and the trajectory-bootstrap 95\% interval
$[0.589,0.714]$.  The interval contains the KPZ value $2/3$ and lies above
the diffusive value $1/2$.
This is a window-defined Fourier length, not a real-space half-width or
the full-chain structure factor.  Its mass constraint is an estimator
convention, not an imposed conservation law of the open chain.
The absolute-amplitude comparison below uses the unadjusted correlations.

The $N=256$ and $N=512$ profiles agree within statistical uncertainty at
$t=12$ [Fig.~\ref{fig:longtime}(c)].  Longer-time size and source-position
comparisons show that boundary sensitivity is concentrated in the tails:
the response in the common central window is substantially less sensitive
than its full second moment (Appendix~\ref{app:deterministic}).
This distinction becomes especially important for the full-chain Fourier
observable, which samples fluctuations next to the reservoirs.

\subsection{Fourier relaxation and the role of boundaries}

\begin{figure}[t]
 \centering
 \includegraphics[width=\columnwidth]{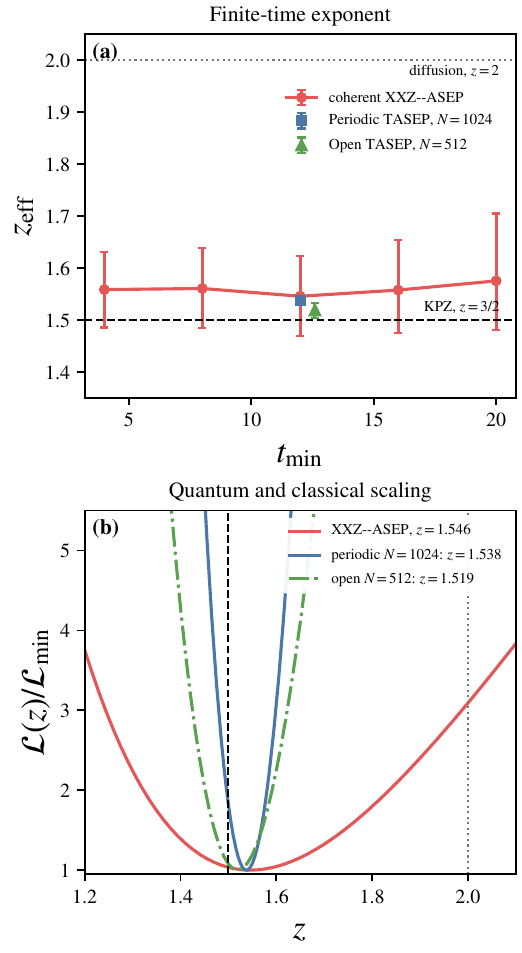}
 \caption{\label{fig:fourier}Finite-time and boundary calibration of the
 Fourier analysis.  (a) Exponent as the lower fit time is varied; the blue
 square is the periodic $N=1024$ reference and the green triangle the
 reservoir-matched $N=512$ reference.  Both reference fits start at
 $t=12$; the green marker is displaced horizontally by $0.6$ for visibility.
 (b) Collapse losses divided by their respective minima.}
\end{figure}

The low-momentum structure factor provides a direct test of the relaxation
time $t_k\sim k^{-z}$.  In the primary window $12\le t\le50$, with
$2\sin(|k|/2)$ between $0.0245$ and $0.196$, the quantum data give
$z=1.546\ [1.469,1.622]$.  The collapse loss at $z=2$ is 2.99 times that
at $z=3/2$.  Varying the lower time cutoff from 4 to 20 leaves the exponent
compatible with KPZ [Fig.~\ref{fig:fourier}(a)].

Finite-time exponents need not equal the asymptotic value even in a known
KPZ process.  The periodic $N=1024$ TASEP reference illustrates this bias,
but does not reproduce the quantum chain's reservoirs.  Matching the
classical reference to the quantum system's $N=512$, boundary rates,
times, momenta, and fitting mask gives
$z_{\rm open}=1.519\ [1.504,1.533]$.  The quantum-minus-classical difference
is $0.027\ [-0.050,0.117]$ and is not statistically resolved.
In contrast, changing only the classical boundary geometry produces a
resolved shift of $-0.020\ [-0.024,-0.017]$.
The reservoir dependence is thus a physical finite-geometry effect, not a
time-step error.  The comparison with the matched classical process
supports the KPZ interpretation within the same finite geometry.
The calibration is described in Appendix~\ref{app:geometry}.

Removing the current asymmetry changes the result.  For $p=q=1$, the same
analysis gives $z=1.924\ [1.756,2.142]$, consistent with diffusion and
separated from the asymmetric case.  Coherent exchange can change the
diffusion coefficient, but the vanishing current curvature removes the
Burgers nonlinearity.  This comparison ties the anomalous spreading to the
asymmetric current rather than to coherent motion alone.

\subsection{The microscopic KPZ amplitude}

\begin{figure*}[t]
 \includegraphics[width=0.98\textwidth]{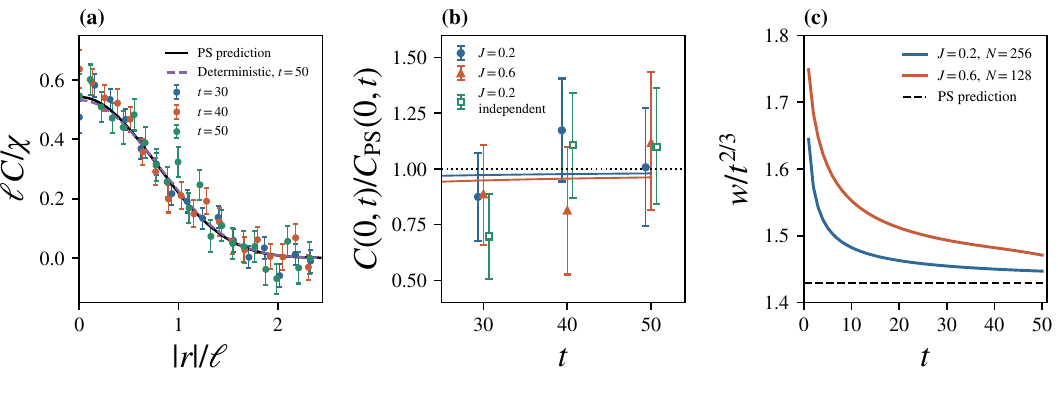}
 \caption{\label{fig:ps}KPZ amplitude fixed by the microscopic current and
 susceptibility, with no fitted length or amplitude.
 (a) Symmetrized production profiles at $J=0.2$, rescaled with
 $\ell=2t^{2/3}$; the solid black line is the Pr\"ahofer--Spohn function.
 The deterministic $t=50$ local response is shown separately.
 Every third spatial point is displayed; bars denote one standard error,
 and all recorded bins enter the analysis.
 (b) Peak divided by its fixed theoretical value.  The independent rerun
 includes the lower $t=30$ peak; bars are trajectory-bootstrap 95\% intervals.
 (c) Deterministic root-mean-square width divided by $t^{2/3}$ over all
 recorded positive times, compared with the leading prediction (dashed).
 The time axis starts at zero; the ratio is undefined at $t=0$, where no
 point is shown.  Local-response curves and source-averaged
 trajectory points are distinct finite-chain estimators.
 Numerical settings and convergence checks are given in
 Appendix~\ref{app:deterministic}.}
\end{figure*}

A more restrictive prediction concerns the profile itself.  The exact
current curvature and susceptibility fix $\ell=2t^{2/3}$ and $\chi=1/4$.
The leading bulk template is
$C_{\rm PS}(r,t)=\chi\ell(t)^{-1}f_{\rm KPZ}[r/\ell(t)]$.
With these values, Fig.~\ref{fig:ps}(a) compares $\ell C/\chi$ against
$r/\ell$ directly with the Pr\"ahofer--Spohn function.  There is no fitted
length, vertical amplitude, background, or peak position.
For the fixed-source response
$C_y(r,t)=\Tr[\delta n_{y+r}e^{t\Lio}(\delta n_y\rho_*)]$,
the full finite-chain RMS width is defined by
\[
 w^2(t)=\frac{\sum_r(r-\bar r(t))^2C_y(r,t)}{\sum_r C_y(r,t)},
 \qquad \bar r(t)=\frac{\sum_r rC_y(r,t)}{\sum_r C_y(r,t)}.
\]
These moments use the unadjusted response over the physical chain at a
fixed source $y$.  The RMS width is $w(t)=\sqrt{w^2(t)}$.
The predicted peak and RMS width follow from
$C_{\rm PS}(0,t)=\chi f_{\rm KPZ}(0)/\ell(t)$ and
$w_{\rm PS}(t)=\sqrt{\mu_2}\,\ell(t)$:
they are
$C(0,t)\simeq0.06781\,t^{-2/3}$ and
$w(t)\simeq1.42902\,t^{2/3}$.

The trajectory profiles are consistent with this leading form in the core
and intermediate spatial range at their statistical precision.  At $t=50$,
the measured-to-predicted peak ratios are $1.008\ [0.742,1.273]$ at
$J=0.2$ and $1.118\ [0.816,1.435]$ at $J=0.6$.
The intervals are broad, so finite-time corrections can be present without
being resolved by these ensembles.  An independent $J=0.2$ ensemble does
resolve a lower peak at $t=30$, about $3.14$ standard errors below the
leading prediction [Fig.~\ref{fig:ps}(b)].  The joint profile diagnostics,
including this deviation, are given in Appendix~\ref{app:amplitude}.

The deterministic responses reveal the size of such corrections more
clearly.  At $J=0.2,N=256,t=50$, the peak is $2.0\%$ below the leading
prediction and the RMS width is $1.3\%$ above it.  These shifts exceed the
observed time-step dependence.  At $J=0.6,N=128$, the corresponding
shifts are $3.8\%$ and $2.9\%$.
Figure~\ref{fig:ps}(c) shows the approach of the scaled width toward its
predicted value over the full positive-time range.  High-statistics TASEP
also has a scaled core mass about $2\%$ below the asymptotic value at
$t=50$.  Resolved deviations at these times are therefore compatible with
an approach to KPZ scaling.  The central profile, its microscopic scale,
and the Fourier relaxation together support that interpretation; the data
do not resolve the far tail or establish the asymptotic regime at every $J$.

\section{Weak-asymmetry crossover}
\label{sec:crossover}

\begin{figure*}[t]
 \includegraphics[width=0.96\textwidth]{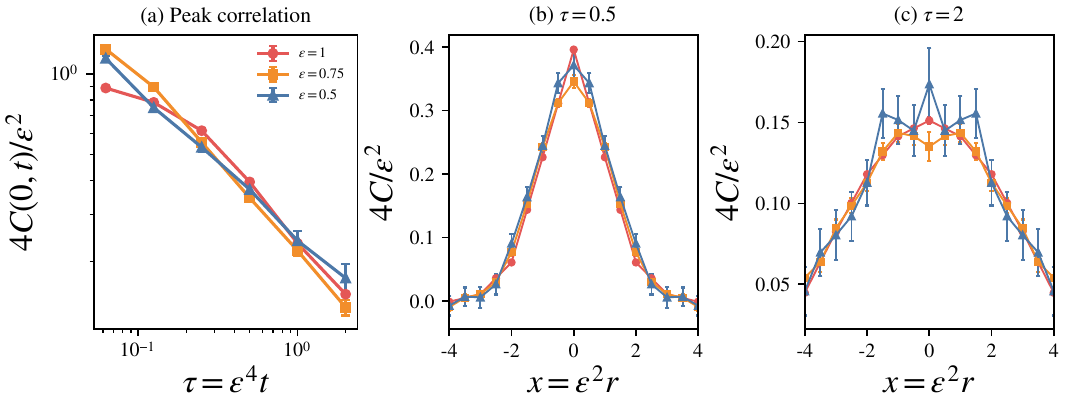}
 \caption{\label{fig:crossover}Edwards--Wilkinson-to-KPZ crossover.
 (a) Central correlations versus $\tau=\epsilon^4t$.
 (b),(c) Profiles at $\tau=0.5$ and $2$
 rescaled according to Eq.~\eqref{eq:crossover-form}.  The collapse improves
 as the common rescaled time increases.}
\end{figure*}

The current curvature also predicts how the onset of KPZ transport moves
as the asymmetry is reduced.  At fixed $p+q=2$, decreasing
$\epsilon=(p-q)/2$ weakens the Burgers nonlinearity and extends the
preasymptotic diffusive regime.  The crossover scales
$\ell_\times\sim\epsilon^{-2}$ and $t_\times\sim\epsilon^{-4}$ therefore
provide a test beyond fitting an exponent at one parameter point.

For $\epsilon=1,0.75,0.5$ on $N=256$, we compare
$4C/\epsilon^2$ at fixed $\tau=\epsilon^4t$ as a function of
$x=\epsilon^2r$.  The profiles differ at early rescaled times, when the
strongly asymmetric chain still probes microscopic relaxation.  Their
agreement improves as $\tau$ increases [Fig.~\ref{fig:crossover}].  On the
common interval $-4\le x\le4$, the diagonal reduced $\chi^2$ decreases
from $3.29$ at $\tau=0.5$ to $1.59$ at $\tau=1$ and $0.60$ at $\tau=2$.
These values compare profiles using pointwise bootstrap variances; shared
trajectory fluctuations prevent interpreting them as independent-bin
probabilities.  The simultaneous length and time rescaling nevertheless
shows the dependence on the microscopic asymmetry expected from nonlinear
fluctuating hydrodynamics.

\section{Discussion}
\label{sec:discussion}

The density dynamics retains the exclusion-process hydrodynamic structure
although its microscopic evolution is not classical.  This is possible
because the stationary current fixes the nonlinear coupling while the
coherences need not supply an additional conserved field.  The jump
algebra constrains exact conservation laws, and the accessible spectrum
identifies the density-visible slow branch.  The real-time profile and
Fourier relaxation then test the consequences of this mode structure.
The diffusion bound of Ref.~\cite{AmpelogiannisDoyon2026} and the gap
scaling in Ref.~\cite{Westhoff2026} address related aspects of the problem;
the present results connect the nonlinear current to the physical density
profile and its microscopic scale.

The stronger-coupling ensemble at $J=0.6$ favors KPZ over diffusive scaling
and agrees within uncertainty with the reservoir-matched TASEP reference.
A separate calculation through $t=100$ finds no resolved drift of the
quantum-to-classical Fourier-scale ratio, although only two low-momentum
points remain reliable at the final time (Appendix~\ref{app:strongj}).
These observations support persistence beyond the main weak-coupling data.
They leave open the dependence of the onset time on $J$: stronger coherent
motion can extend the transient regime, and the $J=1$ results available
here concern short times only.

Charge conservation and nonlinear transport have different physical roles.
The strong $U(1)$ symmetry protects the bulk density, whereas the curvature
$j''(\bar n)$ determines whether its hydrodynamics contains the Burgers
nonlinearity.  If weak charge breaking introduces a uniform relaxation
rate $\gamma_s$, the KPZ regime should be cut off at
$t_s\sim\gamma_s^{-1}$ and $\ell_s\sim\gamma_s^{-2/3}$.
Spin loss and pumping in XXZ chains illustrate the resulting competition
between early transport and dissipative damping \cite{Moca2026}.
By contrast, perturbations that preserve the nonlinear current and do not
introduce another slow field can change the approach to scaling without
changing the hydrodynamic universality class.  The source dependence in
Appendix~\ref{app:sources} is consistent with this picture: several local
charge-conserving perturbations excite the same density branch.

\section{Conclusion and outlook}

A coherent XXZ extension of ASEP provides a direct connection between a
microscopic stationary current and nonlinear density hydrodynamics.
The current remains classical in form while the evolving state develops
quantum coherences.  In the regime studied here, those coherences coexist
with a density-visible slow mode, superdiffusive spreading, KPZ Fourier
relaxation, and the predicted weak-asymmetry crossover.
The current curvature and susceptibility fix the leading length and
amplitude of the profile, making the comparison with the
Pr\"ahofer--Spohn function more restrictive than an exponent fit.

The resolved finite-time corrections and reservoir dependence set the
limits of this conclusion.  The profile test covers the core and
intermediate tail, and the dynamical evidence concerns finite sizes,
times, and coherent couplings.  Determining how the crossover time and
leading corrections vary with $J$ is a natural extension.  A second
direction is to identify which charge-conserving coherent perturbations
preserve this single-density-mode description.  The exact current and
its predicted KPZ scale provide physical benchmarks for both questions.

\section*{Acknowledgments}
This work was supported by the National Natural Science Foundation of China
(Grant No.~12304290 and No.~12674355) and the Beijing National Laboratory for Condensed Matter Physics (Grant No.~2025BNLCMPKF017). L.P. acknowledges support from the Fundamental Research Funds for the Central Universities (010-63263115). Generative AI tools (OpenAI Codex) assisted with manuscript editing and code development.  %The authors take full responsibility for the scientific content.
%\else
%\begin{acknowledgments}
%This work was supported by the National Natural Science Foundation of China
%(Grant No.~12304290) and the Beijing National Laboratory for Condensed Matter
%Physics (Grant No.~2025BNLCMPKF017). L.P. acknowledges support from the Fundamental
%Research Funds for the Central Universities (010-63263115).
%Generative AI tools (OpenAI Codex) assisted with manuscript editing,
%code development, and numerical analysis.  The authors take full
%responsibility for the scientific content.
%\end{acknowledgments}
%\fi

\section*{Data Availability}
The data and analysis code supporting this study are available from the
corresponding author upon reasonable request.

\appendix

\section{Commutant and hydrodynamic mode count}
\label{app:commutant}

Let $X$ commute with $L_j=b_{j+1}^\dagger b_j$ and $L_j^\dagger$ for every
bond.  It then commutes with
\[
 [L_j,L_j^\dagger]=n_{j+1}-n_j.
\]
Within a fixed-$Q$ sector, all $n_j$ can be reconstructed from $Q$ and these
density differences, so $[X,n_j]=0$ and $X$ is diagonal in the occupation
basis.  The condition $[X,L_j]=0$ makes its diagonal entries equal on two
configurations related by an allowed nearest-neighbor particle--hole swap.
For $0<Q<N$ this configuration graph is connected on a periodic chain;
therefore $X$ is proportional to the identity in the sector.  Adding
$H_{\rm XXZ}$ cannot enlarge the commutant.  Within the charge-diagonal observable algebra, the commutant therefore
consists of functions of total charge.  This does not exclude nonlocal
charge-changing operators on the entire operator space: coherences between
the empty and fully occupied states are an exceptional example.  Such
operators lie outside the charge-neutral density-response sector and do
not add a local hydrodynamic density in the finite-density sectors studied
here.  Possible emergent slow modes are addressed separately
by the spectral calculation.

\section{Finite-size and statistical controls}
\label{app:numerics}

\begin{figure}[t]
 \centering
 \includegraphics[width=\columnwidth]{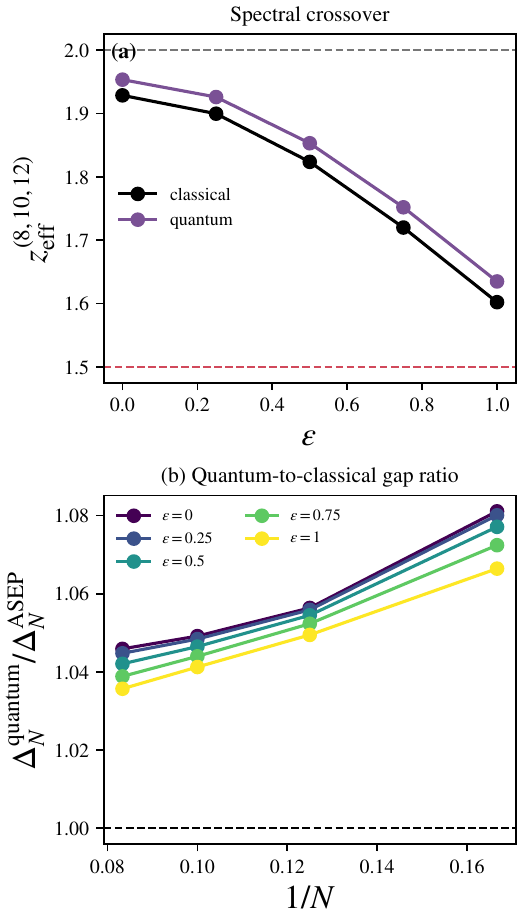}
 \caption{\label{fig:spectral-controls}Finite-size spectral controls.
 (a) Effective exponent from a linear fit of $\log\Delta_N$ against
 $\log N$ at $N=8,10,12$ as the asymmetry $\epsilon$ is increased.  (b) Ratio of the quantum and classical density-branch decay
 rates.  The real-time crossover is examined in the density profiles
 in Fig.~\ref{fig:crossover}.}
\end{figure}

The exact stationary-state and current identities are independently
verified at $N=6$ for all nontrivial charge sectors and
$J=0,0.2,0.6,1$, with residuals below $2\times10^{-16}$.
The spectral calculation reaches $N=14$ in the density momentum sector;
the relative residual of its leading mode is $2.6\times10^{-8}$.
At $J=0$, the off-diagonal decay rate in the half-filled, charge-neutral
sector is 2 for the tested sizes $N=4$--$12$, while the density gap decreases with $N$.
Figure~\ref{fig:spectral-controls} shows the corresponding crossover as
the jump asymmetry is increased: the effective density exponent moves
from the diffusive side toward KPZ, and the quantum-to-classical gap ratio
approaches unity.  These finite-size trends identify the relevant mode;
the real-time correlations supply the larger-system test.

The principal $J=0.2,N=512$ trajectory ensemble uses a bond cap of 64;
the largest observed bond dimension is 31 and the accumulated discarded
weight is below $2.4\times10^{-15}$.  Local channels are checked against
full Lindblad evolution at $N=4$.
A separate $J=0.6$ comparison uses 32 pairs of trajectories with the same
initial seeds at $\delta t=0.05$ and $0.025$.
For real normalized Fourier modes $m=1,\ldots,8$ at
$t=12,20,30,40,50$, the joint centered max-$t$ bootstrap gives $p_{\rm diag}=0.107$.
Here $p_{\rm diag}$ denotes a bootstrap diagnostic value, distinct from the
hopping rate $p$.  Its simultaneous difference bound is too broad to establish time-step
convergence by itself.  Projection onto the fixed PS profile gives the
amplitude shift $\Delta A=A_{0.025}-A_{0.05}=-0.0252\ [-0.1229,0.0743]$, with
$A=\sum C_{\rm PS}C/\sum C_{\rm PS}^2$ and the sums over those five times
and $r=-32,\ldots,32$.  This quantity diagnoses amplitude changes; it is
not applied as a fitted rescaling of the plotted profiles.
The deterministic responses below resolve numerical sensitivity more
precisely.

The central-profile length estimate stabilizes between 256 and 512
trajectories.  A fit to the peak alone is less discriminating,
$\beta_{\rm peak}=0.615\ [0.468,0.794]$.
In the weak-asymmetry comparison, the early rescaled-time peak deviations
are large and decrease before the common profile develops.  Peak-only and
full-profile diagnostics need not vary monotonically together, since they
probe different parts of the finite-time correlation.

\section{Density-response convergence and boundary sensitivity}
\label{app:deterministic}

Direct evolution of $e^{t\Lio}(\delta n_y\rho_*)$ separates resolution
errors from trajectory sampling.  It propagates this traceless, generally nonpositive operator without
imposing normalization.
Comparison with the full Lindbladian at $N=4$ confirms second-order
finite-time convergence.  The operator bond dimension $\chi_{\rm op}$
is distinct from the pure-state trajectory bond dimension.

For $J=0.2$, central-source calculations at $N=128,256$ use
$\delta t=0.05,0.025$, $\chi_{\rm op}=64$, and a relative squared
singular-value cutoff of $10^{-12}$.  At $N=256,t=50$, halving the step
changes the profile by $0.0375\%$ in relative $L^2$ norm and its second
central moment by $0.0723\%$.  Increasing the operator cap to 96 and
tightening the cutoff to $10^{-14}$ at $N=128$ changes the corresponding
quantities by $0.00413\%$ and $0.00849\%$.
At $J=0.6,N=128$, the time-step changes at $t=50$ are $0.0410\%$ and
$0.0847\%$.  A stricter fine-step calculation with
$\chi_{\rm op}=128$ and cutoff $10^{-13}$ changes them by $0.00122\%$
and $0.00352\%$; the maxima over all integer times through $50$ are
$0.00122\%$ and $0.00573\%$.
Relative differences use the finer or stricter result in the denominator.
They measure sensitivity to the stated resolution changes, rather than
rigorous error bounds.

\subsection{Local spreading and reservoir influence}

The full second moment is more sensitive to distant tails than the central
profile.  At $J=0.2,t=50$, increasing $N$ from 128 to 256 changes the
second moment by $0.392\%$.  Moving the source in the $N=256$ chain from
site 65, nearest the reservoir within the sampled source window, to site
128 changes the full profile by $0.0470\%$ and its second moment by
$0.1725\%$.  In the common $|r|\le32$ window, the relative profile
changes for both comparisons are below $1.3\times10^{-7}$.
Profiles on different chains are aligned in displacement; zero extension
is used only outside their physical domains.

At $J=0.6$, independent ensembles of 512 trajectories at $N=256$ and 256
at $N=512$ give no resolved late-time size shift.  The joint profile
comparison over $r=-32,\ldots,32$ and $t=30,40,50$ gives $p_{\rm diag}=0.928$,
but its simultaneous half-widths of $0.0040$--$0.0046$ are comparable to
the central signal.  Matching physical wave numbers in the full-chain
Fourier comparison gives $p_{\rm diag}=0.620$, with half-widths $0.075$--$0.209$.
These statistical intervals leave percent-level size dependence unresolved.

\subsection{Full-chain Fourier response}
\label{app:fourier-controls}

To calculate Eq.~\eqref{eq:full-chain-structure} directly, we evolve the
density perturbations $N^{-1/2}\sum_y\cos(ky)\delta n_y\rho_*$ and
$N^{-1/2}\sum_y\sin(ky)\delta n_y\rho_*$.  Their combined readout includes
all source and observation sites, including the boundaries.
The selected modes are $(N,J,m)=(256,0.6,1)$ and $(512,0.2,2)$,
both at $k=2\pi/256$; the latter is the lowest mode in the primary
$N=512$ collapse.  Each response reaches $t=50$ at time steps $0.05$
and $0.025$ with cutoff $10^{-12}$, followed by a fine-step comparison
at cutoff $10^{-13}$.  The operator caps increase from 96 to 128 for
$J=0.6$ and from 64 to 96 for $J=0.2$.
The changes in Table~\ref{tab:fourier-controls} are small compared with
trajectory sampling uncertainty.  This statement concerns the two
specified modes; it is not a uniform all-momentum bound.

\begin{table}[t]
 \caption{\label{tab:fourier-controls}Direct full-chain Fourier controls.
 Here $\widehat S=4S_N(k,t)$; maxima are over $t=0,1,\ldots,50$.
 Relative changes divide the complex difference by the modulus of the
 finer or stricter result.  Truncation denotes the joint cap-and-cutoff
 change described in the text.}
 \begin{ruledtabular}
 \begin{tabular}{@{}cccc@{}}
 $(N,J,m)$ & Change & $\max|\Delta\widehat S|$ & Relative (\%) \\
 \hline
 $(256,0.6,1)$ & time step & $9.91\times10^{-5}$ & $0.0122$ \\
 $(256,0.6,1)$ & truncation & $1.18\times10^{-5}$ & $0.0015$ \\
 $(512,0.2,2)$ & time step & $1.25\times10^{-4}$ & $0.0147$ \\
 $(512,0.2,2)$ & truncation & $2.61\times10^{-5}$ & $0.0030$ \\
 \end{tabular}
 \end{ruledtabular}
\end{table}

The exact continuous-time full-chain response is real at half filling.
The combined particle--hole transformation and spatial reflection preserves
the generator and $\rho_*$ and sends
$\delta n_x\to-\delta n_{N+1-x}$.  The two density signs cancel, while
reflection conjugates the Fourier phase.  Since the real-space correlation
is real, $S_N(k,t)=S_N(k,t)^*$.
Finite time steps, truncation, and sampling can leave imaginary residuals.
A joint real-plus-imaginary comparison with the deterministic response at
$t=12,20,30,40,50$ gives $p_{\rm diag}=0.058$ for the main $J=0.6$ ensemble and
$p_{\rm diag}=0.242$ for the main $J=0.2$ ensemble.  An independent $J=0.6$ ensemble
at $t=30,40,50$ gives $p_{\rm diag}=0.020$, dominated by its $t=50$ imaginary
component, $2.83$ standard errors from the prediction.
This residual remains part of the comparison.  These are exploratory
consistency diagnostics; the principal exponent fits use the real part.

\subsection{Geometry-matched classical reference}
\label{app:geometry}

For each of $N=256,512$, the classical reference contains 16384 paired
open and periodic TASEP trajectories, with half-filled Bernoulli initial
states and open-chain injection and extraction rates one.
The continuous-time process is sampled exactly, without a classical time
step.  Pairs share initial configurations and stochastic proposals, while the
two sizes use disjoint seeds.  Both geometries use the quantum observation
grid $t=0,12,16,20,30,40,50$ and its momenta.
The fitting mask is fixed by
$0.08<\operatorname{Re}S_N(k,t)/S_N(k,0)<0.98$ in the quantum ensemble
mean and is held fixed in all comparisons.
Confidence intervals use 2048 whole-trajectory bootstrap resamples, with
classical pairs drawn together and quantum trajectories independently.
Normalization follows ensemble averaging.

\begin{table}[t]
 \caption{\label{tab:geometry}Fourier calibration with matching finite
 geometry.  The last column gives the quantum-minus-open effective
 exponent with its bootstrap 95\% interval.}
 \begin{ruledtabular}
 \begin{tabular}{ccc}
 $(N,J)$ & $z_{\rm open}$ & $z_{\rm quantum}-z_{\rm open}$ \\
 \hline
  $(256,0.6)$ & $1.494$ & $0.015\ [-0.087,0.127]$ \\
 $(512,0.2)$ & $1.519$ & $0.027\ [-0.050,0.117]$ \\
 \end{tabular}
 \end{ruledtabular}
\end{table}

The open-minus-periodic exponent shifts at the same size are
$-0.047\ [-0.054,-0.040]$ for $N=256$ and
$-0.020\ [-0.024,-0.017]$ for $N=512$.
At $t=50,k=2\pi/256$, the normalized structure-factor shifts are
$-0.0726$ and $-0.0364$, respectively.  They demonstrate a measurable
reservoir effect.  Matching this geometry leaves the quantum-to-classical
exponent difference unresolved (Table~\ref{tab:geometry}), without implying
identical finite-time corrections or a thermodynamic-limit result.

\section{Absolute-profile diagnostics}
\label{app:amplitude}

We evaluate $f_{\rm KPZ}(s)=g_{\rm PS}''(s)/4$, where $g_{\rm PS}$ is the
stationary height-covariance scaling function of
Ref.~\cite{PraehoferSpohn2004}, using its published table and a derivative-based
interpolation.  Its integrated mass
and second moment reproduce the convention used in Eq.~\eqref{eq:kpz-amplitude}.
Every theoretical feature is evaluated on the same integer spatial grid as
the corresponding data, without forcing the observed mass to $\chi$.

The joint amplitude diagnostic comprises the peak and cumulative masses
within $|r|/\ell\le0.5,1,1.5$ at $t=30,40,50$.
Complete trajectories are resampled jointly across these features; a
centered maximum standardized residual supplies the familywise diagnostic.
The main $J=0.2$ and $J=0.6$ values are $p_{\rm diag}=0.605$ and $0.394$,
respectively.  The $J=0.6,N=256$ profiles cover $|r|\le32$; at
$t=40,50$ this does not cover $|r|/\ell\le1.5$, so those two features
are excluded rather than padded with zeros.  An independent
$J=0.2$ ensemble gives $p_{\rm diag}=0.0107$, dominated by its $t=30$ peak;
the independent $J=0.6$ ensembles at $N=256,512$ give $p_{\rm diag}=0.510,0.566$.
Independent ensembles are analyzed separately.

A complementary shape-only test fits one core length against a periodic
$16384$-trajectory TASEP reference at $t=128$, after imposing the central-window
mass convention described above.  Its outer-window diagonal
reduced $\chi^2=0.769$ over $t=12,20,30,40,50$ and resolved signal to
$|s_{\rm fit}|\simeq2.1$, where $s_{\rm fit}=r/\ell_{\rm fit}(t)$, refer to
that fitted convention.  This complementary test
does not fix the absolute amplitude; its finite-time reference is not the
exact Pr\"ahofer--Spohn function.

\section{Dependence on the local source}
\label{app:sources}

The Hermitian two-site operators commuting with the local charge form a
six-dimensional real vector space.  At half filling a basis is
$\{\1,\delta n_j+\delta n_{j+1},\delta n_j-\delta n_{j+1},
Z_j,X_j,Y_j\}$, with $Z_j=\delta n_j\delta n_{j+1}$,
$X_j=a_j+a_j^\dagger$, $Y_j=i(a_j-a_j^\dagger)$, and
$a_j=b_{j+1}^\dagger b_j$.
Here $Y_j$ is a unit-amplitude bond source; its sign and normalization give
the coherent current $-JY_j$, and it differs from the total current in
Eq.~\eqref{eq:current-op}.
The identity has no connected response, the density sum excites the
hydrodynamic mode, and the density difference adds a lattice derivative.
For $O=n,X,Y,Z$, define $O_k=\sum_j e^{ikj}O_j$, with $n_j$ replaced by
$\delta n_j$ for the density source.  The quoted amplitude ratio is the
late-time ratio of
$\Tr[n_k^\dagger e^{t\Lio}(O_k\rho_M)]$ to the same response with $O_k=n_k$.
All sources use this convention without separate norm normalization.
The quoted amplitudes are moduli of these generally complex ratios.
ED gives $|A_Y/A_n|$ decreasing from 0.344 at $N=6$ to 0.125 at $N=12$;
a fit over these sizes gives an effective $k^{1.46}$ dependence, without
establishing its asymptotic power.  The bond-coherence ratio remains below
$0.4\%$.  The connected pair source has the same late-time decay rate as
the density; at $N=12$ its amplitude ratio has modulus 0.08100 in the coherent
model and 0.08135 in TASEP (Table~\ref{tab:sources}).

\begin{table}[t]
\caption{Dependence on the local source.  Amplitudes are normalized by the
density slow-mode amplitude; the connected-pair values are quoted at $N=12$.}
\label{tab:sources}
\small
\begin{tabular}{@{}ll@{}}
\toprule
Source & Slow-mode diagnostic \\
\midrule
Bond coherence $X$ & $\max|A_X/A_n|=3.63\times10^{-3}$ \\
Bond source $Y$ & $0.344\ (N=6)\to0.125\ (N=12)$; $\propto k^{1.46}$ \\
Connected pair $Z$ & $0.08100$ (quantum); $0.08135$ (TASEP) \\
\bottomrule
\end{tabular}
\end{table}

\section{Dependence on the coherent coupling}
\label{app:strongj}

\begin{figure*}[t]
 \includegraphics[width=0.96\textwidth]{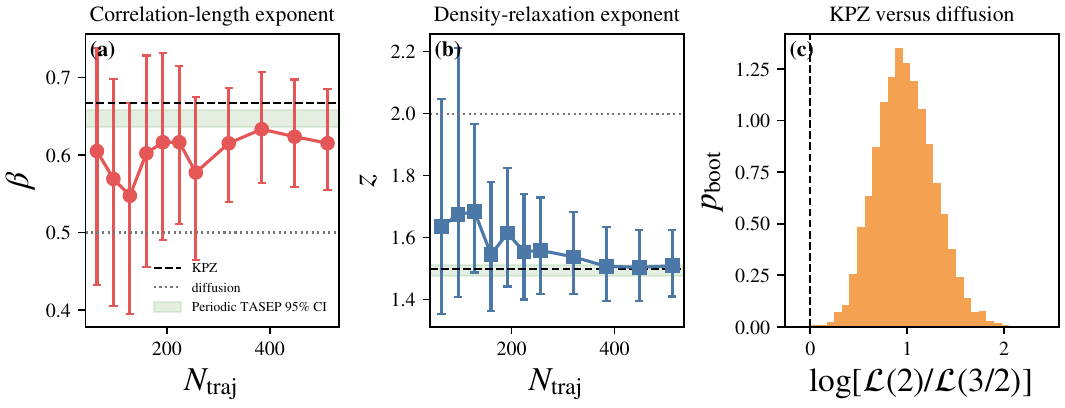}
 \caption{\label{fig:strongj}Independent high-statistics test at $J=0.6$.
 The $N=256$ ensemble contains 512 trajectories through $t=50$.
 (a) Full-chain Fourier half-height-length exponent $\beta_S$
 (denoted $\beta$ in the plot) and (b) low-momentum collapse exponent versus cumulative
 trajectory count.  The shaded TASEP 95\% intervals use the periodic width reference in (a)
 and the reservoir-matched open reference in (b).
 (c) Bootstrap distribution of the log loss ratio between fixed diffusive and
 KPZ exponents; positive values favor KPZ, as occurs in 99.96\% of the
 bootstrap samples.  Error bars are trajectory-bootstrap 95\% intervals.}
\end{figure*}

\begin{table}[t]
\caption{Additional controls at $N=256$ and $J=0.6$.  Parentheses give sample
counts and brackets give trajectory-bootstrap 95\% intervals.  Wave-number arguments are suppressed in the Fourier ratio.  The fixed
template projection $A$ is defined in Appendix~\ref{app:numerics}.
These small ensembles do not determine the primary
exponent.}
\label{tab:j06-controls}
\small
\begin{tabular}{@{}ll@{}}
\toprule
Control & Outcome \\
\midrule
$\delta t=0.05/0.025$ (32) &
$\Delta A=-0.0252\ [-0.1229,0.0743]$ \\
$\chi_{\max}=64/96$ (32) &
$\max|\Delta[S_N(t)/S_N(0)]|=4.3\times10^{-11}$ \\
$t_{\max}=100$ (32) &
$R(100)=0.961\ [0.777,1.221]$ \\
\bottomrule
\end{tabular}
\end{table}

To examine the range of the weak-coupling result, we simulated $J=0.6$ on
$N=256$ with 512 independent trajectories through $t=50$.  In the matched
low-momentum window the collapse exponent is
\[
 z_{J=0.6}=1.509\ [1.410,1.625],
\]
while the geometry-matched open TASEP reference gives
$z_{\rm open}=1.494\ [1.475,1.513]$.  Their difference,
$0.015\ [-0.087,0.127]$, is unresolved.  The periodic reference
gives $z=1.538\ [1.528,1.547]$ but does not match the reservoirs.
The collapse loss
at $z=2$ is 4.27 times that at $z=3/2$; trajectory bootstrap samples favor the
KPZ value in 99.96\% of resamples.  This fraction measures resampling
stability, not a posterior probability for the universality class.
For the full-chain observable, $\ell_S(t)=1/\widehat k_{1/2}(t)$ uses the
first interpolated half-height crossing of the running-minimum envelope of
$\operatorname{Re}S_N(k,t)/S_N(k,0)$ on the sampled lattice-wave-number grid.
Fitting $\ell_S\propto t^{\beta_S}$ gives
$\beta_S=0.615\ [0.555,0.685]$, which contains $2/3$.  After applying the same
finite-time estimator to the periodic TASEP reference, the width-exponent difference is
$-0.032\ [-0.096,0.039]$.  As a collapse-independent check on the relaxation time,
fixed-level crossings of Fourier modes give
$z=1.529\ [1.279,1.786]$; the corresponding diffusive-to-KPZ loss ratio is
27.3.

The dependence on size and sampling is consistent with this interpretation.  An $N=128$ ensemble is
statistically compatible with the $N=256$ profiles (familywise max-$t$
$p_{\rm diag}=0.581$).  A 32-seed paired bond-cap comparison between $\chi_{\max}=64$ and 96
changes the ensemble-normalized structure factor by at most
$4.3\times10^{-11}$ on the sampled grid.  Splitting the 512 trajectories into
eight consecutive cohorts shows no excess heterogeneity in the collapse
exponent ($p_{\rm diag}=0.435$), and the exponents from the two halves differ by
$-0.081\ [-0.296,0.131]$.  Finally, a separate 32-trajectory pilot through $t=100$ tests the
microscopic scale using an empirical periodic TASEP reference.
Let $\Phi_T(v)$ denote its normalized Fourier profile at $t=128$,
with argument $v=k w_T(128)$ and $w_T$ the reference RMS displacement
about $r=0$, evaluated in its saved central window.
At each pilot time we fit a dilation $R(t)$ through
$S_N(k,t)/S_N(k,0)\simeq\Phi_T[kR(t)w_{\rm PS}(t)]$.
Here $k=2\pi m/N$ is the physical wave number and
$w_{\rm PS}(t)=\sqrt{\mu_2}\,\ell(t)$, so $R$ is a length ratio,
not a relaxation-rate ratio.  The geometric mean over pilot times is
$R_{\rm global}=0.992\ [0.874,1.117]$, and
$R(100)=0.961\ [0.777,1.221]$.
The one-component prediction is $R=1$ in the bulk scaling limit.
This pilot fits $R$ against a finite-time reference; it is distinct from the
absolute, unadjusted PS comparison in Fig.~\ref{fig:ps}.
Since the $t=100$ estimate uses only two reliable
low-momentum points, it is a check against a gross late-time drift rather than
an independent asymptotic-exponent measurement.  The three additional
controls are summarized in Table~\ref{tab:j06-controls}.  Thus the
stronger-coupling data independently favor KPZ over diffusive scaling in the
observed window, although they cannot establish the far-tail scaling function
or the behavior at arbitrary $J$.

At $J=1$, an $N=128$ paired comparison of MPS bond caps 64 and 96 gives a
profile RMS difference of $3.02\times10^{-7}$ and reduced $\chi^2=1.15$ at
$t=12$.  Some trajectories reach the larger cap, so this is only a short-time
convergence check and does not support an asymptotic claim.

\end{document}